# ELECTRON MOBILITY IN DENSE ARGON GAS AT SEVERAL TEMPERATURES

**A.F.Borghesani**

Istituto Nazionale per la Fisica della Materia

Department of Physics, University of Padua,

Via F. Marzolo 8, I-35131 Padua, Italy

**P. Lamp**[1]

Max-Planck-Institut fuer Physik

Werner-Heisenberg-Institut

Foehringer Ring 6, D-80805 Munich, Germany

---

[1] Present address: BMW AG, D-80788 Munich, Germany

**ABSTRACT**

The mobility $\mu$ of excess electrons in dense Argon gas has been measured as a function of the applied electric field $E$ and of the gas density $N$ at several temperatures in the range *142.6<T<200 K*, encompassing the critical temperature $T_c$=*150.86 K*. We report here measurements at densities up to $N \approx 7\ nm^{-3}$, close to the critical density, $N_c \approx 8.1\ nm^{-3}$, reached for the isotherm closest to the critical one. At all temperatures, below as well as above $T_c$, and up to moderately high densities, the density-normalized mobility $\mu N$ shows the usual electric field dependence present in a gas with a Ramsauer-Townsend minimum due to the mainly attractive electron-atom interaction. $\mu N$ is constant and field independent for small values of $E$, shows a maximum for a reduced field $E/N \approx 4\ mTd$, and then decreases rapidly with the field, approximately proportional to $(E/N)^{-1/2}$. The zero-field density-normalized mobility $\mu_0 N$, for all $T>T_c$, shows the well known anomalous positive density effect, i.e., $\mu_0 N$ increases with increasing $N$, confirming previous results obtained for *T=300 K* and *162.7 K*. Below $T_c$, however, $\mu_0 N$ does not show the expected anomalous positive density effect, but it rather features a broad maximum. This appears to be a crossover behavior between the positive density effect shown for $T>T_c$ and the small negative effect previously observed for $T \approx 90\ K$. In any case, the data at all temperatures confirm the interpretation of the anomalous density effect as being essentially due by the density-dependent quantum shift of the electron ground state kinetic energy in a disordered medium as a result of multiple scattering (MS) processes, although other MS processes do influence the outcome of the experiment.

## 1. INTRODUCTION

The mobility µ of excess electrons in dense non-polar gases is a well suited probe to investigate the nature of the electronic states in a disordered system. In a dense gas the transport properties of electrons depend in a complicated way on the interplay between the dense environment and the quantum nature of the electron.

The usual picture of classical kinetic theory, i.e., that of binary electron-atom collisions well separated in space and time, breaks down because of the short interatomic distance at high density and because of the fairly large extension of the electron wavepacket, especially at low temperature. As the gas density is increased, possibly at low temperature, the electron mean free path gets shorter and shorter and it eventually becomes comparable in magnitude to the electron thermal wavelength. Moreover, the long thermal wavelength of the electrons makes them interact simultaneously with more than one atom at once. In these conditions, multiple scattering (MS) processes influence the electron energetics and, hence, its transport properties.

The classical kinetic theory for the electron mobility [1] predicts that the so called zero-field density-normalized mobility $\mu_0 N$, at constant temperature $T$, does not depend on the gas density $N$, and depends on the type of the gas only through the electron-atom momentum transfer scattering cross section, $\sigma_{mt}$, that embodies the quantum nature of the electron-atom interaction. According to that theory, the density-normalized mobility for near thermal electrons is given by

$$(\mu_0 N)_{cl} = \frac{4}{3} \frac{e}{(2\pi m)^{1/2} (k_B T)^{5/2}} \int_0^\infty \frac{\varepsilon}{\sigma_{mt}(\varepsilon)} e^{-\varepsilon/k_B T} d\varepsilon \qquad (1)$$

where *m* and *e* are the electron mass and charge, respectively, and $k_B$ is the Boltzmann's constant. Eq. (1) shows that, once the type of the host atoms is known, the density-normalized mobility for near thermal electrons depends only on the gas temperature *T*, but not on the gas density *N*.

However, the experiments have shown quite early that, even in the noble gases, the simplest possible systems, and even at moderate density, the mobility significantly deviates from the classical prediction, Eq. (1). Anomalous density effects have been experimentally observed in a number of compressed gases [2-13]. A *negative density effect*, i.e., the density-normalized mobility at thermal energies, $\mu_0 N$, decreases with incresing *N* at constant *T* (and eventually drops rapidly to very low values because of the formation of localized electron states), is shown by gases, such as He and Ne, whose interaction with the excess electrons is dominated by the short-range repulsive exchange forces. These repulsive gases are endowed with a positive scattering length *a*.

On the contrary, a *positive density effect*, i.e., $\mu_0 N$ increases with increasing *N*, is shown by gases, such as Ar, whose scattering length *a* is negative because of the predominant long-range, attractive polarization electron-atom interaction. In Argon, and in the heavier noble gases, the electron mobility remains high also in the liquid and in the crystalline solid phases [14,15], therefore leaning support to the idea that electrons in these gases remain in extended states.

The experimentally observed anomalous density effects have been interpreted in terms of the appearance, at high density, of the effects of multiple scattering as a consequence of two major facts: the mean free path $\ell$ of electrons becomes comparable to its thermal wavelength $\lambda_T$ and the electron wavepacket spans over a region so large as to contain in it several (even thousands) scatterers.

Several multiple scattering theories have been developed, in the limit of vanishingly small electric fields, to rationalize the experimental observations [17-19]. They are all based on the presence of a density-dependent quantum shift of the electron energy in a dense disordered medium [20-22], and, for repulsive gases like Neon and He, introduce a mobility edge. In their present form, the theories disagree more or less with the experimental data, although the main idea of the energy shift is now well accepted.

However, the results and the analysis of accurate measurements in Ne [10,11] and in Ar [12,13] (but also in liquid Ar [16]), two gases, whose momentum transfer scattering cross sections, though for different physical reasons, depend strongly on the electron energy, have shown that the different behavior of the mobility in repulsive and attractive gases can be rationalized in an unified picture, if all multiple scattering effects are taken into account heuristically.

The heuristic model, henceforth referred to as BSL [12], has identified three main multiple scattering effects that are simultaneously at work to modify the measured mobility with respect to the classical prediction:

a) The density-dependent quantum shift of the electron kinetic energy;
b) The correlation among scatterers, which is very important in the proximity of the critical point [23];
c) An enhancement of the backscattering rate due to the quantum self-interference of the electron wavepacket along paths connected by time-reversal symmetry [24,25].

With the choice of the proper parameters for each gas, namely the appropriate cross section and equation of state, the heuristic BSL model does describe accurately both the positive and negative density effect on the mobility.

This model strongly relies on the existence of the density-dependent quantum shift of the ground state energy of excess electrons in the dense medium. However, in contrast to the MS theories, it assumes that only the kinetic part of the shift affects the mobility in the same way for both attractive and repulsive gases, whereas the MS theories introduce different mechanisms of action according to the sign of the total energy shift [17-19, 25].

An other unsolved question in Ar is whether the anomalous density effect remains positive at all temperatures. In fact, there are measurements in a restricted density range for *T ≈ 90 K* showing that the density effect has changed to negative [26].

We have therefore carried out further mobility measurements in Ar gas in a wide density and temperature range in order to get an experimental assessment of these two issues. We report here mobility measurements in Ar gas as a function of the gas density at three temperatures above $T_c$ (*T=152.6, 177.3,* and *199.7 K*) up to a maximum density of *N ≈ 7 nm$^{-3}$* for the isotherm closest to the critical one. We also report measurements at a temperature (*T=142.6 K*) below $T_c$ up to a density *N ≈ 3.2 nm$^{-3}$*.

## 2. EXPERIMENTAL DETAILS

We have used the well known pulsed photoemission method as in our previous measurements in Ne gas [10] and we have exploited the same experimental apparatus used for electron drift mobility measurements in liquid and critical Ar [14]. For details we refer to previous papers. We want to recall here only some important experimental features.

The production of photoelectrons is accomplished by irradiating a gold photocathode with a *1-µs* short UV-pulse of a Xe-flashlamp (Hamamatsu model L2435). The lamp repetition rate has been chosen low, 2 – 5 *Hz*, in order to avoid pile up of $O_2^-$ ions in the drift space. The current induced by the drifting electrons is integrated by means of a charge-sensitive amplifier, whose output is connected to a digital storage oscilloscope. Each signal acquisition is sent over the GPIB bus to a Personal Computer. For each temperature and field strength settings, approximately *256* signals were software-averaged to improve the signal-to-noise ratio. The analysis of the signal in order to determine the electron time-of-flight has been carried out by exploiting the mathematical methods developed for the Ne measurements [27]. The induced signal waveform of the electron drifting at constant speed in a very pure gas sample is a straight line, and the drift time is easily determined by the analysis of the waveform. The overall accuracy of the mobility measurements is $|\Delta\mu/\mu|$ 5 %.

In order to reduce the influence of $O_2^-$ impurities on both amplitude and shape of the signal, particularly severe at low field strengths where the drift time is large, great care has been devoted to purify the gas. We used ultra-high purity Ar "60" (99.9999 % by vol.), further purified by an Oxisorb filter. We also used high-temperature bakeable stainless steel tubings and valves. The impurity concentration is estimated to be in the p.p.b. range [28].

The gas pressure was measured by means of a Heise Bourdon gauge (model 710B). The cell in the cryostat was thermoregulated at the desired temperature within *±1 mK*. The temperature is measured by means of a calibrated PT-1001 platinum resistor (Lakeshore CryotronicsInc.). Pressure readings were converted to density values by using the equation of state given by Tegeler *et al.* [29]. The estimated error in the calculated density values is < 0.5 %.

### 3. **EXPERIMENTAL RESULTS AND DISCUSSION**

We have carried out measurements at four temperatures, namely, *T=199.7*, *177.3*, and *152.6 K* above the critical temperature, and *T=142.6 K*, below it. The explored density range depends on the temperature, since the experimental cell can withstand pressures up to ≈ *6.0 MPa*. Therefore, the highest values of the density, $N \approx 7\ nm^{-3}$, has been reached at the temperature closest to the critical one.

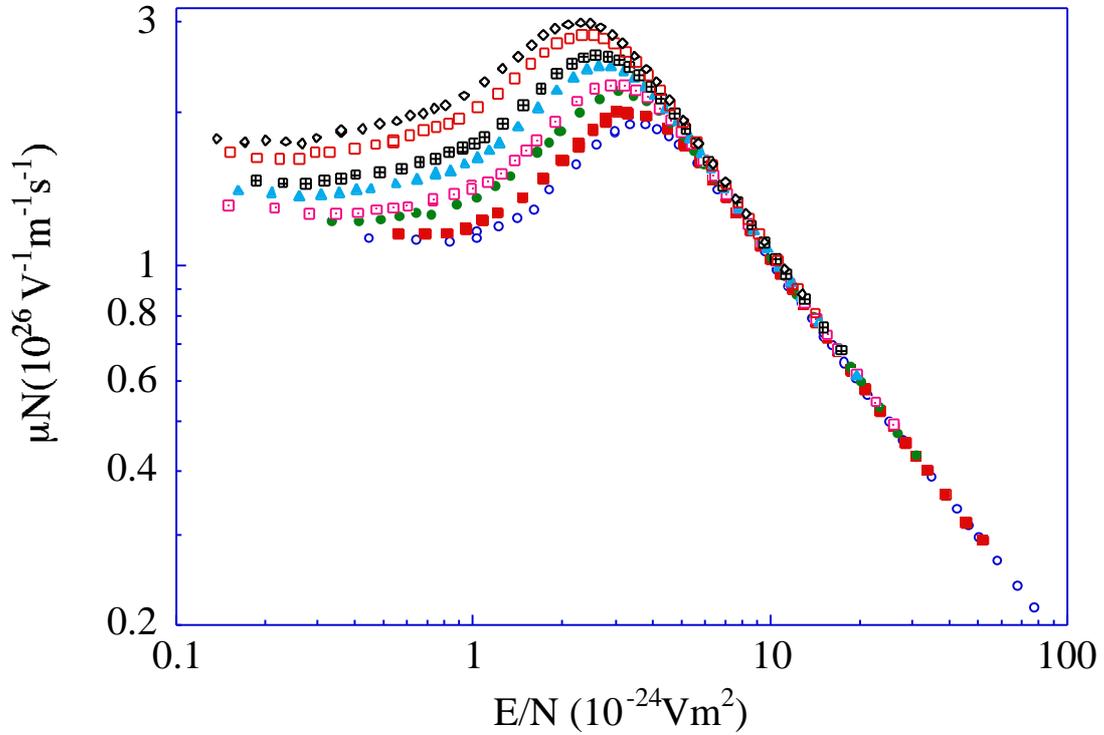

**Figure 1.** Density-normalized mobility $\mu N$ as a function of the reduced electric field *E/N* at *T≈200 K* for several densities (in units of $nm^{-3}$) *N* = 0.512 (open circles), 0.768 (closed squares), *1.282* (closed circles), *1.541* (dotted squares), *2.058* (closed triangles), *2.312* (crossed squares), *2.888* (open squares), *3.133* (open diamonds).

For all densities and temperatures we have measured the drift mobilities as a function of the electric field strength *E*. It is customary to introduce the reduced electric field *E/N* because it is related to the energy that electrons gain from the field. We have investigated a three-orders-of-magnitude range in *E/N*, *0.1<E/N<100 mTd* (*1 mTd = $10^{-24}$ Vm$^2$*). This range corresponds to electron energies from a few *meV* (the thermal energy around *T=150 K* amounts to ≈ *20 meV*) up to a few (*2-3*) *eV*.

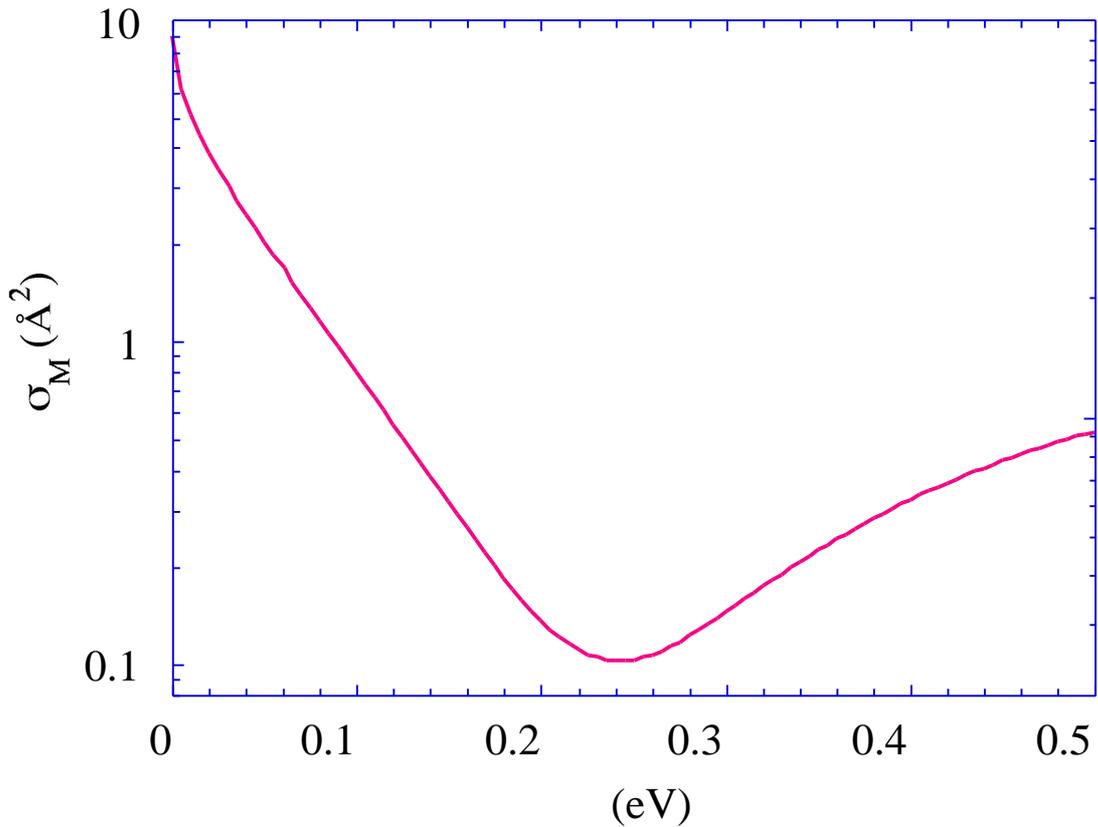

**Figure 2.** Electron-atom momentum transfer scattering cross section for Ar showing the Ramsauer-Townsend minimum at an electron energy of about *0.24 eV*.

The field dependence of μ*N* shows the features typical of a gas whose scattering cross section has a Ramsauer-Townsend minimum. By focusing the attention on the curve of the lowest density in Figure 1, we note that μ*N* is nearly constant at small fields. Then, it shows a maximum for a reduced field $(E/N)_{max} \approx 4$ *mTd*. Finally, $\mu N$ decreases approximately as $(E/N)^{-1/2}$ for large values of *E/N*. The mobility maximum is due to the Ramsauer-Townsend (RT) minimum of the momentum transfer scattering cross section, shown in Figure 2 [30,31]. The RT minimum is located at an electron energy of $\varepsilon_{RT}$ *230-240 meV*. The mobility maximum appears at a value of the reduced field *E/N* such that the average electron energy $<\varepsilon>$ equals $\varepsilon_{RT}$.

By inspecting Figure 1, again, it is easily observed that the values of $\mu N$ at small fields increase with increasing density *N*. This is the well known *positive density effect*. We also note that the position of the mobility maximum $(E/N)_{max}$ clearly shifts to smaller values as the gas density increases. Since the maximum appears at the lowest densities, it must be related to the RT minimum of the cross section even at higher densities. Its shift to smaller values when *N* increases means that at higher densities less energy is required from the field for the mean electron energy to reach $\varepsilon_{RT}$. In some sense, as it will be specified later on, the environment, through the density, appears to supply an increasing, positive contribution to the mean electron energy.

A very striking, experimental confirmation that the mean electron energy has a positive contribution from the density and that the density produces on the mean electron energy an effect equivalent to that due to the temperature is shown in Figures 3 and 4.

In Figure 3 we show the density-normalized mobility $\mu N$ as a function of the reduced field for several temperatures at fixed density $N \approx 2.5$ *nm$^{-3}$*. Literature data [12] at an intermediate temperature are also shown for comparison.

At constant density, the temperature produces several important effects on $\mu N$. On one hand, the zero-field value, $\mu_0 N$, increases with increasing *T*. This effect is a consequence of the combined effect of the increase of the mean electron energy with the temperature and of the energy dependence of the cross section shown in Figure 2. For energies below the RT minimum $\sigma_{mt}$ decreases rapidly with energy and therefore the zero-field mobility, which can be considered as a suitable average of the inverse cross section (see Eq. 1), increases with temperature. (We note that at high field the dependence on the temperature fades away and all curves drop onto a single one. This is because at high fields the thermal contribution to the mean electron energy is negligible with respect to the energy contributed by the electric field.)

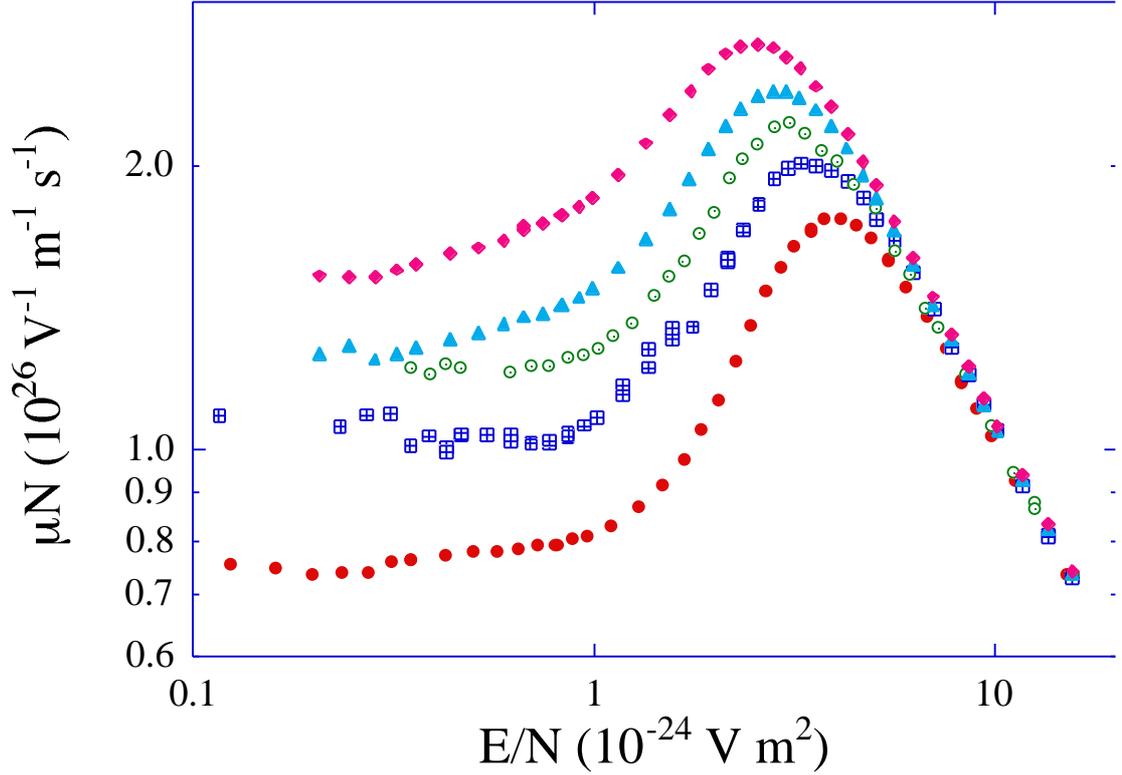

**Figure 3.** Density-normalized mobility $\mu N$ as a function of the reduced field $E/N$ at fixed density $N \approx 2.5\ nm^{-3}$ for several temperatures $T =$ *142.6 K* (closed circles), *152.6 K* (crossed squares), *162.7* (dotted circles, ref. [12]), *177.3 K* (closed triangles), and *199.7 K* (closed diamonds).

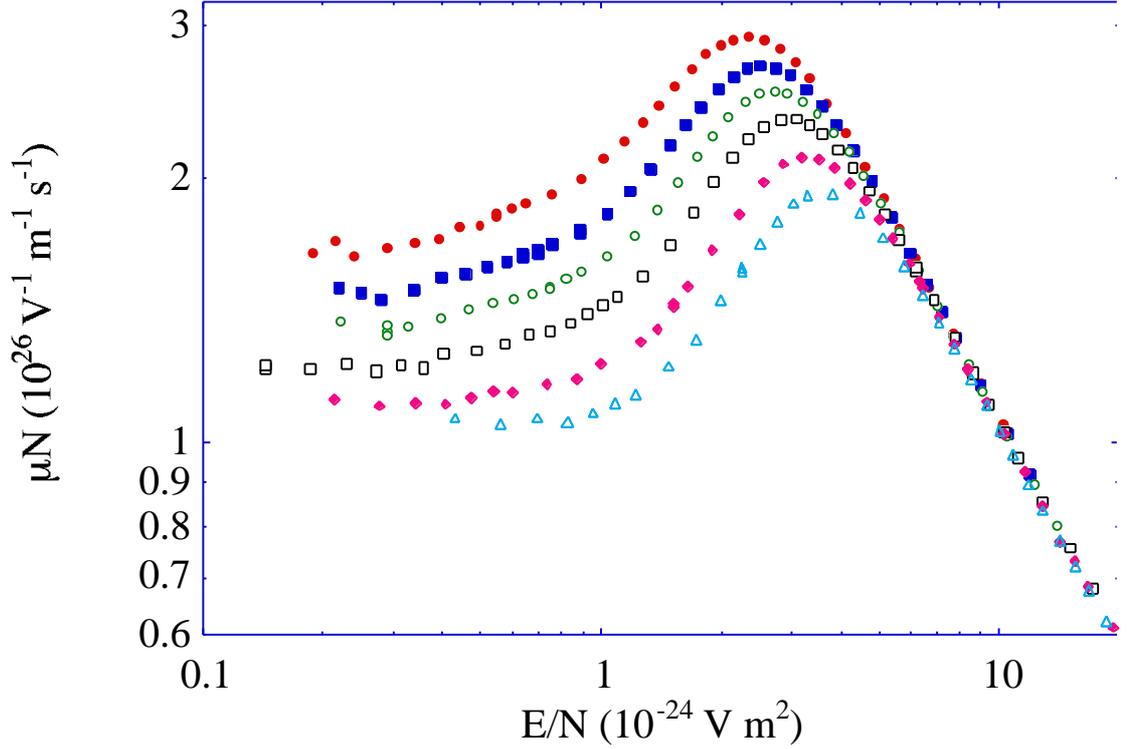

**Figure 4.** Density-normalized mobility $\mu N$ as a function of the reduced field $E/N$ at constant $T=177.3\ K$ for several densities $N$ (in units of $nm^{-3}$). $N=3.865$ (closed circles), $3.321$ (closed squares), $2.835$ (open circles), $2.312$ (open squares), $1.542$ (closed diamonds), and $0.767$ (open triangles).

On the other hand, we observe also that the position of the mobility maximum, $(E/N)_{max}$, in Figure 3 shifts to smaller values as the temperature increases. Again, this observation is rationalized by noting that, at the mobility maximum, $<\varepsilon>\approx \varepsilon_{RT}$, and that the increase of $T$ increases the thermal contribution of the mean electron energy so that less energy is required from the field for the mean energy to reach the value corresponding to the RT minimum.

We now show, in Figure 4, $\mu N$ as a function of $E/N$ at constant temperature $T=177.3\ K$ for some values of the density and compare these data with those reported in

Figure 3. With increasing density, at constant *T*, the zero-field values of μ*N* increase, the position of the mobility maximum shifts to smaller values, and all curve merge into a single one at large *E/N* values. In other words, the behavior of μ*N* as a function of *E/N* is affected in much the same way by either temperature or density.

In some sense, *T* and *N* can be interchanged as far as their effect on the electron mobility is concerned. The obvious conclusion to be drawn from the comparison of Figures 3 and 4 is that the density increases the average kinetic energy of the electrons by a positive contribution since the increase of temperature does the same.

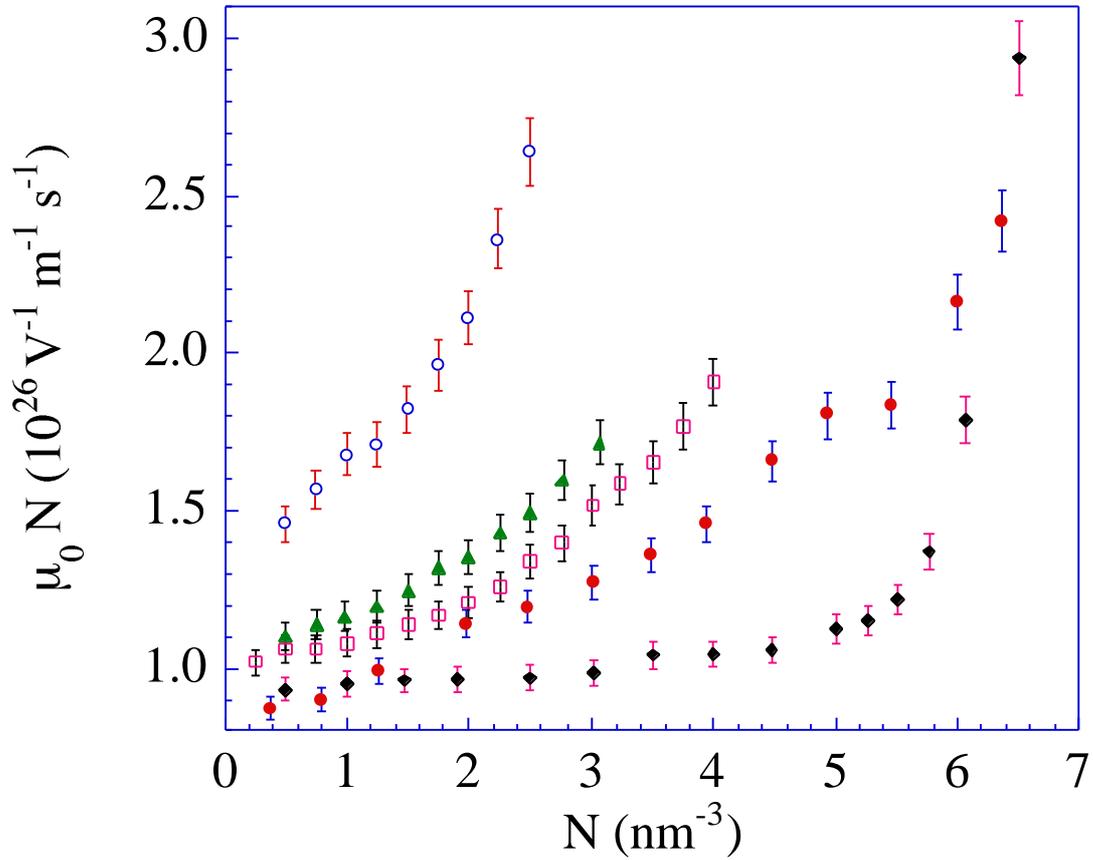

**Figure 5.** Zero-field density normalized mobility $\mu_0 N$ as a function of the density *N* for several temperatures. *T =298 K* (open circles, ref. [3]), *199.7 K* (closed triangles), *177.3 K* (open squares), *162.7 K* (closed circles, ref. [12]), and *152.6 K* (closed diamonds).

The same conclusion, i.e., that the density provides a positive contribution to the mean electron kinetic energy, is drawn by inspecting the density dependence of the mobility at near thermal energies $\mu_0 N$, shown in Figure 5 for the three supercritical isotherms. In this Figure we also show literature data for comparison. Our $\mu_0 N$ data at *T=199.7, 177.3*, and *152.6 K* fit well with previous results and show the expected increase with *N*. We note that, up to moderately high values of density, $N \leq 6$ *nm$^{-3}$*, the rate of increase of $\mu_0 N$ with *N* decreases when the temperature is lowered toward the critical temperature. The *152.6 K* isotherm represents a special case because here we have been able to reach quite large density values, $N \approx 7$ *nm$^{-3}$*. Further new measurements at this temperature up to $N\approx9.5$ *nm$^{-3}$* are presently under scrutiny but they apparently confirm the present results and, in particular, the concept of the density-dependent energy shift. These new data and the relative discussion, however, will be presented in detail in a subsequent publication.

The positive density effect shown by the near thermal density-normalized mobility in Figure 5 can be rationalized by introducing the above mentioned three MS effects in the framework of kinetic theory according to the BSL model [12]. The first effect is the density-dependent quantum shift of the electron ground-state kinetic energy. Clues for this effect have been shown previously. According to MS theories [20-22], the ground state energy of an electron in a dense medium is shifted by a quantity $V_0(N)$ that, for Ar, is negative. According to calculations of Springett *et al.* [32], $V_0(N)$ can be written as

$$V_0(N) = U_P(N) + E_k(N) \qquad (2)$$

$U_P$ is a potential energy contribution arising from the screened polarization interaction of the electron with the atoms of the environment. $E_k$ is a kinetic energy contribution, due essentially to excluded volume effects and from the short-range translational symmetry of the medium. Owing to its nature of kinetic energy, $E_k$ is positive and increases with increasing $N$, since the volume accessible to electrons shrinks. This kinetic energy contribution has to be added to the usual electron kinetic energy when the scattering properties, namely, the cross sections, have to be calculated.

The second MS effect is due to the correlation among the scatterers, described by the static structure factor of the gas, $S(k)$, where $k$ is the transferred momentum. At low temperatures, the thermal wavelength of the electrons is so large as to encompass a region containing several atoms. Close to the critical point, this number can be of the order of several hundreds or thousands of atoms. The electron wavepacket is then scattered simultaneously by many scatterers and the total scattered wavepacket is obtained by summing up coherently all partial scattering amplitudes contributed by each atom [23].

The third MS effect is an enhancement of the backscattering rate due to the quantum self-interference of the electron wavepacket scattered off atoms located along paths connected by a symmetry operation of time-reversal [24]. This phenomenon has its analogue in the so called *weak localization regime* of electronic conduction in disordered solids, precursor of the Anderson localization [33]. This effect depends on the ratio between the electron wavelength and its mean free path ($\lambda_T/\ell$). For Ar, in the condition of the experiment, ($\lambda_T/\ell$)= $N\sigma_{mt}\lambda_T <1$, and to first order the cross section is enhanced by the factor $(1+ N\sigma_{mt}\lambda_T/\pi)$ [25].

The density-normalized mobility, in the density-modified kinetic BSL model [12], is then obtained by inserting these MS effects in a heuristic way into the equation of classical kinetic theory. We start with the usual equation for $\mu N$ [1]

$$\mu N = -\frac{e}{3}\left(\frac{2}{m}\right)^{1/2}\int_0^\infty \frac{\varepsilon}{\sigma_{mt}(\varepsilon + E_k)}\frac{dg}{d\varepsilon}d\varepsilon \tag{3}$$

$g(\varepsilon)$ is the Davydov-Pidduck electron energy distribution function [34]

$$g(\varepsilon) = A\exp\left[-\int_0^\varepsilon \left(k_BT + \frac{M}{6mz}\left(\frac{eE}{N\sigma_{mt}(w)}\right)^2\right)^{-1}dz\right] \tag{4}$$

$M$ is the atomic mass. $g$ is normalized as $\int_0^\infty z^{1/2}g(z)dz = 1$.

The previously mentioned MS effects can be taken into account by suitably dressing the atomic cross section in the following way

$$\sigma_{mt}(w) = \mathcal{F}(w)\sigma_{mt}(w)\left[1 + 2\hbar N\frac{\mathcal{F}(w)\sigma_{mt}(w)}{(2mw)^{1/2}}\right] \tag{5}$$

where $w = \varepsilon + E_k(N)$ is the shifted electron energy. The cross section is evaluated at the shifted energy. The second term in the parentheses describes the effect of the quantum self-interference, while $\mathcal{F}(w)$ is the Lekner factor that takes into account the correlation among scatterers [23]

$$\mathcal{F}(k) = \frac{1}{4k^4}\int_0^{2k}q^3 S(q)dq \tag{6}$$

where $k$ is related to energy by the usual relation $\varepsilon = \hbar^2 k^2/2m$. $S(q)$ is the static structure factor of the gas, which can be expressed to a good approximation in the Ornstein-Zernicke form as

$$S(q) = \frac{S(0) + (qL)^2}{1 + (qL)^2} \tag{7}$$

$S(0) = Nk_BT\chi_T$ is the long-wavelength limit of the static structure factor and $\chi_T$ is the isothermal compressibility of the gas. $L$ is the correlation length defined as $L^2 = 0.1\, l^2[S(0)-1]$. $l \approx 5\text{-}10$ Å is the short-range correlation length [34].

The shift of the electron kinetic energy can be calculated according to the Wigner-Seitz (WS) model [36] as $E_k(N) = E_{WS} = \hbar^2 k_0^2 / 2m$ where the wavevector $k_0$ is obtained self-consistently as the solution of the eigenvalue equation

$$\tan[k_0(r_S - \tilde{a}(k_0))] - k_0 r_S = 0 \tag{8}$$

where $r_S = (3/4\pi N)^{1/3}$ is the radius of the WS sphere and $\tilde{a}$ is the hard-core radius of the Hartree-Fock potential for Ar. We estimate it from the total scattering cross section as $\tilde{a} = \sqrt{\sigma_{tot}/4\pi}$ [32].

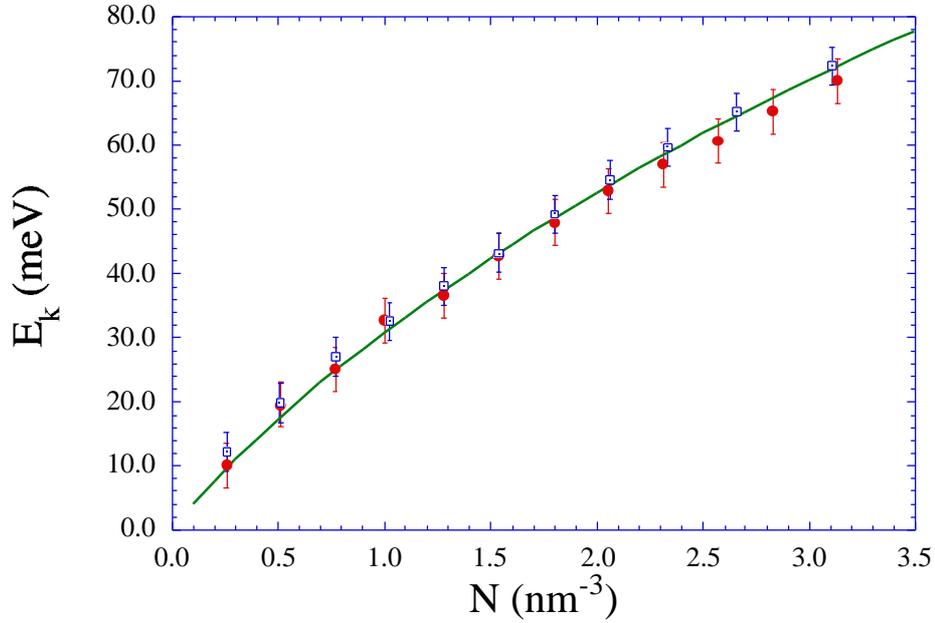

**Figure 6.** Kinetic contribution $E_k$ of the quantum shift of the ground state energy of excess electrons as a function of the Argon gas density $N$ as determined from the mobility data for two different temperatures, $T=199.7\ K$ (closed circles) and $142.6\ K$ (dotted squares). The solid line is the prediction of the Wigner-Seitz model.

The procedure of the analysis of the experimental data with the previous equations consists in adjusting the value of $E_k$ until the calculated zero-field mobility agrees with the experimental value within 1%. We therefore show in Figure 6 the comparison between the values of $E_k (N)$ determined by this fitting procedure and the prediction of the WS model, Eq. (8).

The comparison is shown only for two temperatures, namely $T=142.6$ and $199.7 K$. The agreement between the experimentally determined data and the model is very satisfactory. The results for all other temperatures agree similarly well up to highest densities investigated on each isotherm.

These results confirm the great success of this model obtained for Ar at $T=162.7 K$ [12] and $T=152.15 K$ up to densities $N \leq 7 nm^{-3}$ [13]. For larger $N$, it has been shown that a positive $E_k (N)$ makes the model reproduce the experimental $\mu_0 N$ data very well but that it deviates significantly from the WS model for $N>7 nm^{-3}$. It has been therefore suggested that the use of the WS model at such high densities might not be appropriate anymore [12,13,16].

The second question addressed in the introduction is whether the anomalous density effect remains positive at all temperatures, even below the critical one. The question arises from the experimental observations [26] that at $T=90 K$ $\mu_0 N$ decreases with increasing $N$ for pressures up to $10^5 MPa$.

In order to answer this question we have carried out measurements at $T=142.6 K$, up to a density $N \approx 3.2 nm^{-3}$. In Figure 7 we report our experimental results for $\mu_0 N$. While the electric field dependence of $\mu N$, though not shown here, is similar to that shown in Figure 1, including the maximum related to the RT minimum of the cross section and its shift to smaller field strength as $N$ is increased, its behavior at near thermal energies as a function of the density is quite different from that of the isotherms above $T_c$.

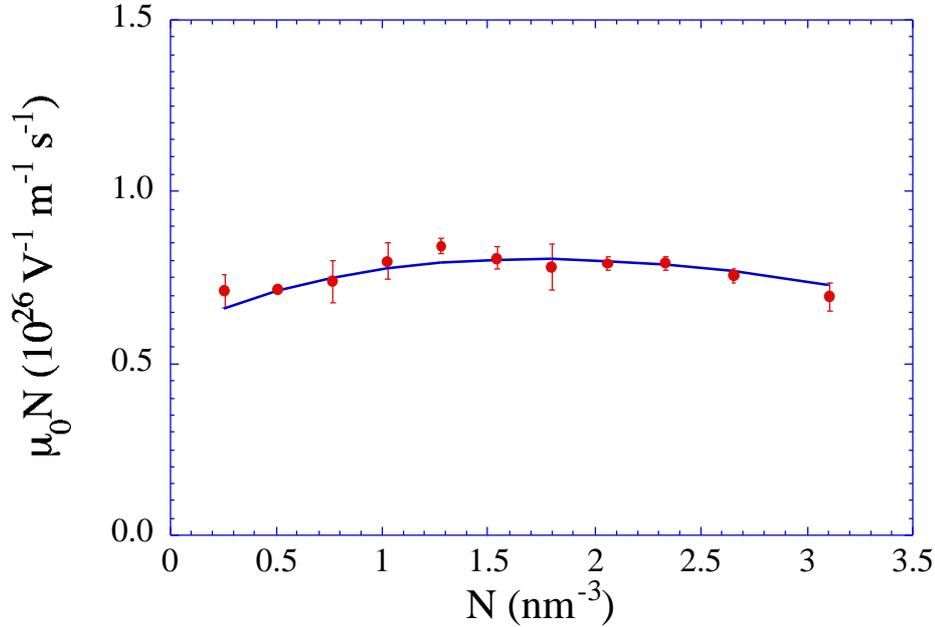

**Figure 7.** Zero-field density-normalized mobility $\mu_0 N$ as a function of the density $N$ at constant $T=142.6$ K, below the critical temperature ($T_c=150.86$ K). The solid line is the prediction of the heuristic BSL model.

At this temperature, the anomalous density effect on $\mu_0 N$ is neither positive nor negative. For small densities, up to $N \approx 1.3$ $nm^{-3}$, $\mu_0 N$ increases with $N$, but it shows a broad maximum, approximately located in the range $(1<N<2)$ $nm^{-3}$ and decreases, though slightly, with increasing $N$ for larger $N$. This is the only other experimental evidence that the anomalous density effect in Ar is not only positive as acknowledged until now. There is apparently a crossover between the density dependence of $\mu_0 N$ at temperatures well above and well below $T_c$. The density effect changes from positive above $T_c$ to negative well below. In any case, the heuristic BSL model describes accurately this unexpected density dependence of $\mu_0 N$, as it is shown in Figure 7. The solid line in the Figure is obtained by inserting the kinetic energy shift $E_k(N)$, as calculated within the WS model, Eq. (8).

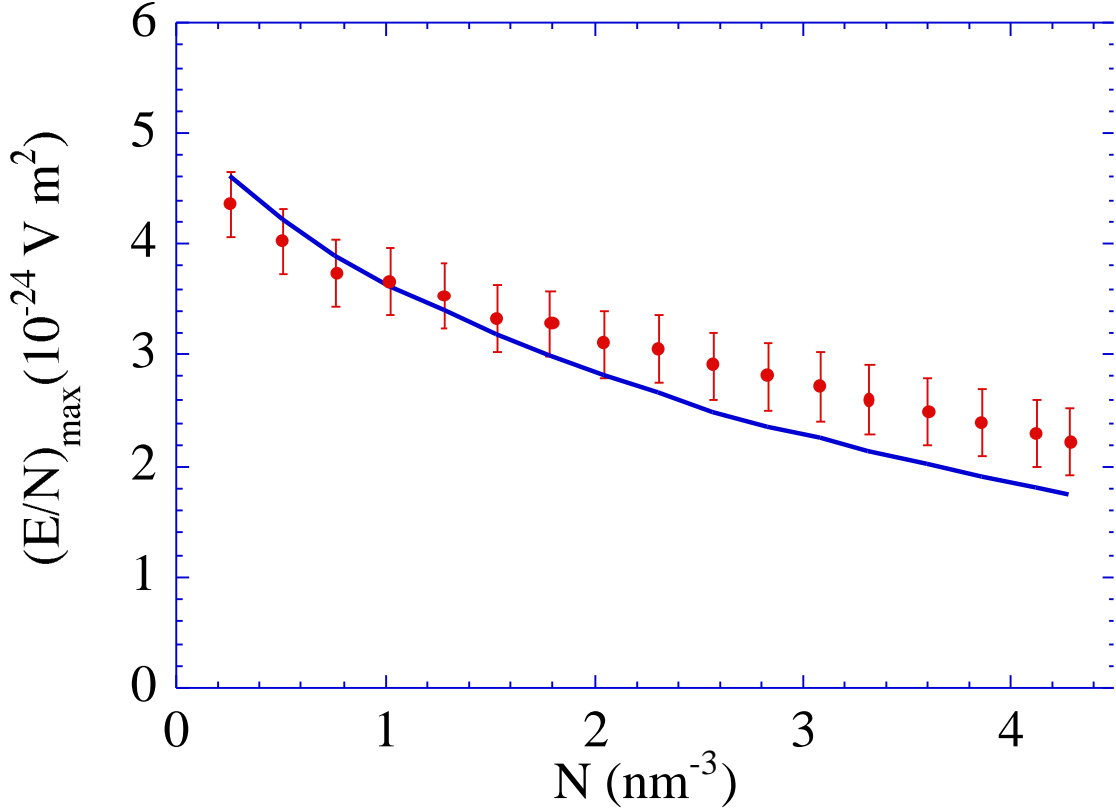

**Figure 8.** Density dependence of the reduced field at the mobility maximum, *(E/N)$_{max}$*, for *T=177.3 K*. The circles are the experimentally determined data, the solid line is the prediction of the BSL model based on the WS calculation of the quantum shift of the electron ground state kinetic energy. Similar plots are obtained at all temperatures.

As a final evidence that the BSL model, with the strong hypotheses made on the energy shift and on the other MS effects, does accurately describe all of the experimental features, we show in Figure 8 the decrease of the reduced electric field at the mobility maximum, $(E/N)_{max}$, as a function of the gas density. The data reported in the figure refer to the *177.3 K* isotherm, but similar results are obtained for all temperatures, above as well as below $T_c$. The solid line is obtained by computing the full field dependence of $\mu N$ for each density according to the equations of the BSL model and then locating the maximum in the computed curve. Again, the agreement between theory and experiment is very satisfactory. This picture confirms the idea that less energy is required from the electric field, when the density increases, for the mean electron energy to become equal to the energy of the RT minimum. More explicitly, this picture once again tells that there is a positive, density-dependent shift of the electron kinetic energy and that this shift is satisfactorily computed from the WS model, at least up to densities $N \approx 7\ nm^{-3}$.

## 5. CONCLUSIONS

The new mobility data presented here, measured at temperatures well above, close to, and well below $T_c$, confirm the results obtained previously in Ar gas. The interpretation of the anomalous density effects arisen from the previous measurement in dense Neon and Argon gases receives strong support from these new measurements. The positive and negative density effects on the electron mobility in dense gases can now be described within an unified picture, where all MS and quantum effects are taken into account heuristically. It is interesting to note that these measurements confirm also the fact that in Ar gas the anomalous density effect does even change sign upon lowering the temperature below the critical one. This confutes the old idea that the anomalous density effect always has a direction opposite to the sign of the electron-atom scattering length.